\documentclass[sigconf]{acmart}

\usepackage[ngerman,english]{babel}
\usepackage{subfig}
\usepackage[binary-units=true]{siunitx}

\usepackage{amsmath}

\DeclareSIUnit\kilok{k}

\AtBeginDocument{%
	\providecommand\BibTeX{{%
			\normalfont B\kern-0.5em{\scshape i\kern-0.25em b}\kern-0.8em\TeX}}}

\copyrightyear{2022}
\acmYear{2022}
\setcopyright{rightsretained}
\acmConference[NICE 2022]{Neuro-Inspired Computational Elements Conference}{March 28-April 1, 2022}{Virtual Event, USA}
\acmBooktitle{Neuro-Inspired Computational Elements Conference (NICE 2022), March 28-April 1, 2022, Virtual Event, USA}
\acmDOI{10.1145/3517343.3517376}
\acmISBN{978-1-4503-9559-5/22/03}

\begin{document}

	\title{Demonstrating BrainScaleS-2 Inter-Chip Pulse-Communication using \mbox{EXTOLL}}

	\author{Tobias Thommes, Sven Bordukat, Andreas Grübl, Vitali Karasenko, Eric Müller, Johannes Schemmel}
	\email{thommes@kip.uni-heidelberg.de}

	\affiliation{%
		\institution{Kirchhoff-Institute for Physics}
		\streetaddress{INF 227}
		\postcode{69120}
		\city{Heidelberg}
		\country{Germany}
	}

	\renewcommand{\shortauthors}{Thommes, et al.}

	\begin{abstract}
		The BrainScaleS\nobreakdash-2 (BSS\nobreakdash-2) Neuromorphic Computing System currently consists of multiple single-chip setups, which are connected to a compute cluster via Gigabit-Ethernet network technology.
		This is convenient for small experiments, where the neural networks fit into a single chip.
		When modeling networks of larger size, neurons have to be connected across chip boundaries.
		We implement these connections for BSS\nobreakdash-2 using the \mbox{EXTOLL} networking technology.
		This provides high bandwidths and low latencies, as well as high message rates.
		Here, we describe the targeted pulse-routing implementation and required extensions to the BSS\nobreakdash-2 software stack.
		We as well demonstrate feed-forward pulse-routing on BSS\nobreakdash-2 using a scaled-down version without temporal merging.
	\end{abstract}



	\maketitle

	\section{Introduction}

\begin{figure}[b]
		\centering
		\includegraphics[width=0.95\linewidth]{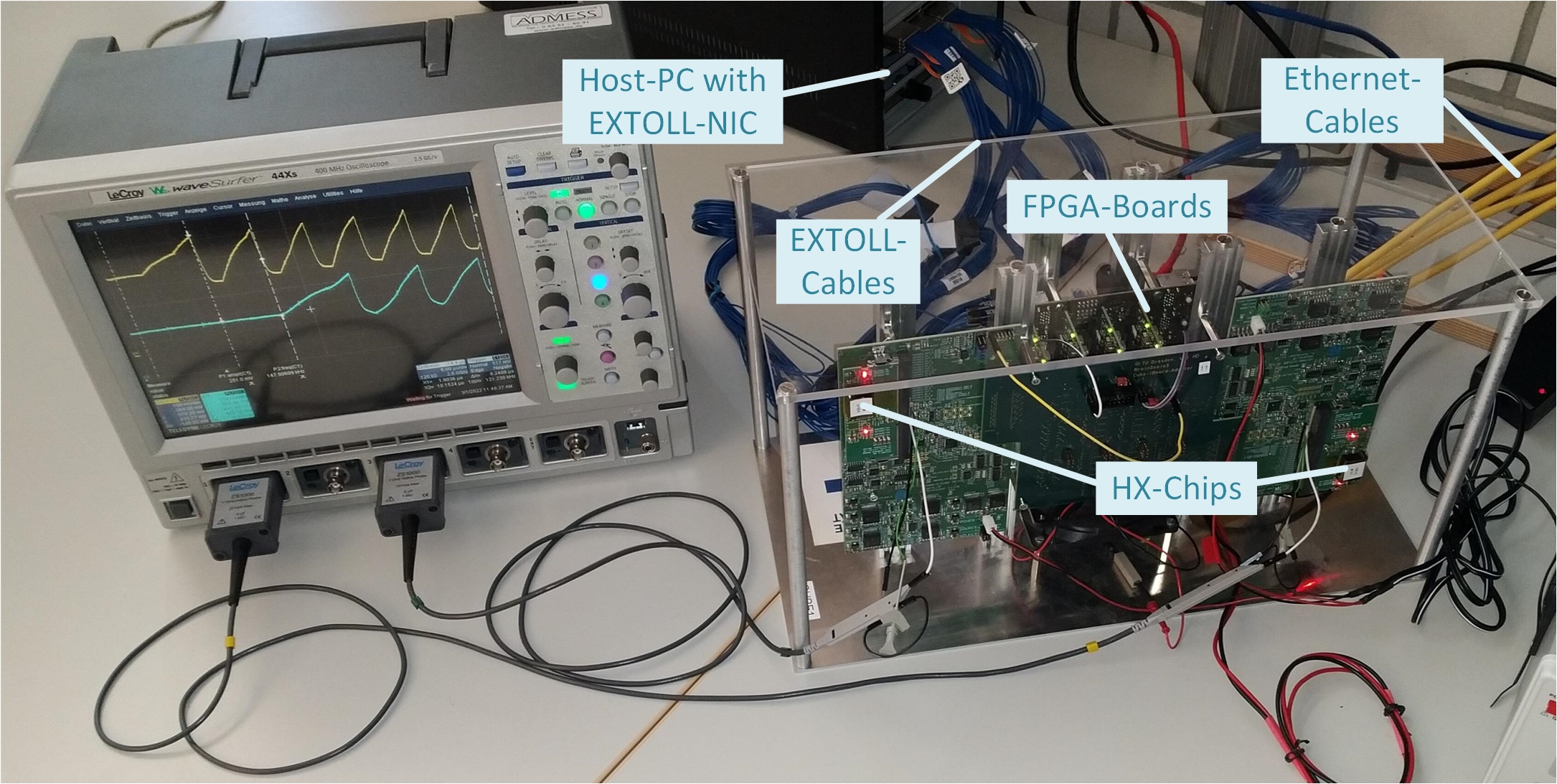}
		\caption{BSS\nobreakdash-2 Lab Setup connected to \mbox{EXTOLL} network. Membrane-traces on two connected HX-chips are shown.}
		\label{fig:CubePhoto}
	\end{figure}

	The \mbox{EXTOLL} network technology \cite{nussle2009fpga, nussle2009rma, froning2013rates, litz2008velo} is based on the Tourmalet Network Interface Card (NIC).
	It offers \num{7} links and implements all the switching and interfacing capabilities, necessary to build an HPC network.
	Each \mbox{EXTOLL} link can comprise up to \num{12} lanes of \SI{8.4}{\giga\bit\per\second} each.
	The NIC can be connected to a host computer through its PCIe~x16~Gen3 connector.
	In an \mbox{EXTOLL} network, the nodes are usually, but not necessarily connected in a 3D-Torus topology, which offers good scaling characteristics.
	Routing of messages through the network is done by the Tourmalet chips and is based on the \SI{16}{\bit} destination node-address.\\
	BSS\nobreakdash-2 as a mixed-signal neuromorphic computing system is built upon the HICANN\nobreakdash-X (HX) chip
	which features \num{512} adaptive exponential integrate and fire (AdEx) neuron-circuits and $\num{512} \times \num{256} = \num{131072}$ synapses \cite{mueller2020bss2ll}.
	Up to \SI{16}{\kilok} synaptic inputs per neuron are configurable by combining neuron circuits. Realizing large networks with such neurons requires a multi-chip system.
	\cite{pehle2022brainscales2_nopreprint, gruebl2020verification, schemmel2020accelerated, billaudelle2020versatile}\\
	Recently the BSS\nobreakdash-2 system development advanced to a multi-chip system featuring \num{46} HX chips,
	each connected to a Kintex~7 \mbox{FPGA} through \num{8}~\SI{1}{\giga\bit\per\second} serial links.
	These systems make use of the BSS\nobreakdash-1 wafer module infrastructure,
	imitating a full wafer-scale implementation by placing many chips on a large PCB of the exact same size and pin-configuration as a BSS\nobreakdash-1 wafer \cite{schemmel2010iscas, schmitt2017hwitl}.
	We consider the topology described in \cite{thommes2021bssextoll} to be optimal for interconnecting multiple \mbox{FPGAs} on wafer modules regarding bandwidth and network diameter.\\
	Figure~\ref{fig:CubePhoto} shows the current lab setup for testing the BSS\nobreakdash-2 \mbox{EXTOLL} networking \cite{schemmel_iscas2012, mueller2020bss2ll}. It is physically connected to the \mbox{EXTOLL} network via USB~3.0 plugs attached to the \mbox{FPGAs'} MGT-ports. Additionally it is still connected to an Ethernet network for \mbox{FPGA} bitfile flashing purposes. This setup contains four \mbox{FPGAs} and two chips.
	
	\begin{figure*}[t]
	\centering
	\includegraphics[width=0.9\linewidth]{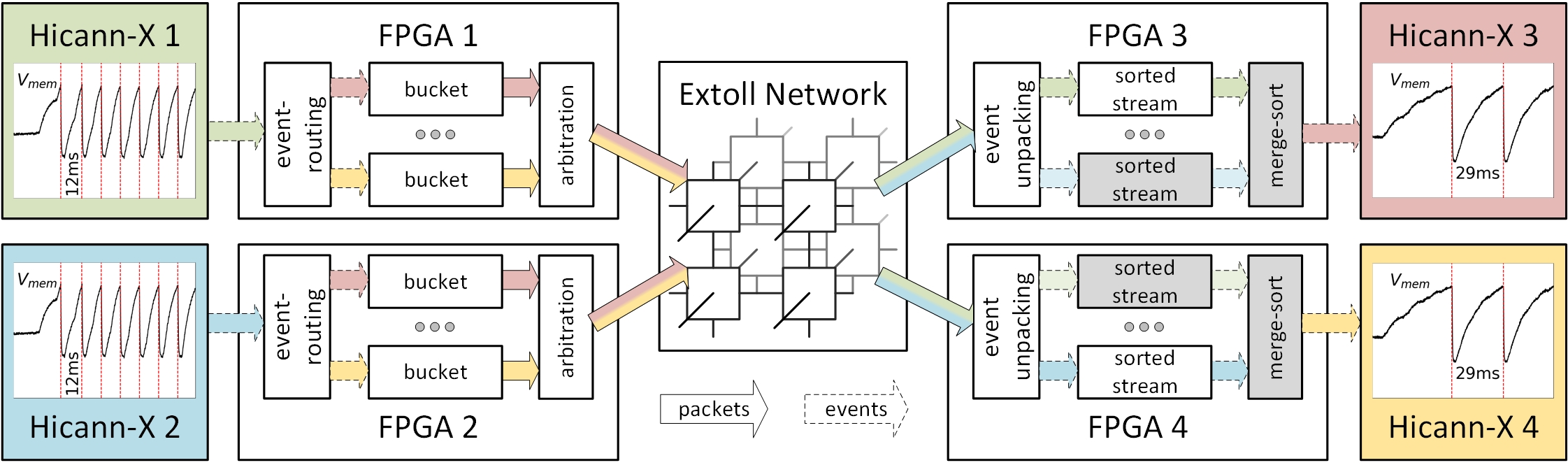}
	\caption{Experiment Setup for an inter-chip feed forward neural network. The Inter-Spike-Interval (ISI), given in bio-time units, is higher at target neurons, as they do not fire with each received input-event.}
	\label{fig:ExperimentSetup}
	\end{figure*}
	
	\section{Host Communication}

	In order to integrate the \mbox{EXTOLL} network with the existing BSS\nobreakdash-2 software stack \cite{mueller2021scalable}, we use the custom protocol layer \textit{Neuromorphic Hardware Transport Layer for \mbox{EXTOLL}} (\textit{\mbox{NHTL EXTOLL}}). This layer sits on top of the \mbox{EXTOLL} network's API for Remote Direct Memory Access (RDMA), \textit{librma2} \cite{nussle2009rma}, and beneath the \mbox{FPGA} software interface (\textit{hxcomm}) \cite{mueller2021scalable}. Thereby, the existing experiment-flow can transparently use the \mbox{EXTOLL} network.

	\subsection{The EXTOLL Protocol}

	The \mbox{EXTOLL} network chip uses RDMA to facilitate low-latency communication using its Remote Memory Access (RMA) unit. The RMA unit consists of three sub-units, the Requester, Responder, and Completer. These handle the different aspects of each RDMA \textit{put} or \textit{get} operation. Messages can be issued with flags such that the sub-units produce notifications upon forwarding them. \cite{nussle2009rma}
	These notifications are used to ensure that API commands are properly executed. Additionally, as they can carry small amounts of payload data, we use notification packets to synchronize the \mbox{FPGA's} send queue with a ring buffer in the host memory, where the \mbox{FPGA} sends its data using RMA-messages.

	\subsection{Interaction with the BSS Software Stack}

	For integration with the BSS\nobreakdash-2 software stack, \textit{\mbox{NHTL EXTOLL}} provides two main functionalities:
	First, it creates and manages the necessary buffers on the host node and configures the \mbox{FPGAs} using the \mbox{EXTOLL} network's Remote Registerfile Access (RRA) feature.
	Second, it provides wrapper functions for receiving and sending data via RDMA with the same syntax used by the higher level components of the software stack.
	With this, the \mbox{EXTOLL} network can be used without having to touch the pre-existing infrastructure provided by the higher abstraction levels \cite{mueller2020bss2ll}.

	\section{Inter-FPGA Communication}
	
	The pulse-communication architecture described for the BSS\nobreakdash-1 system in \cite{thommes2021bssextoll} can easily be adapted for BSS\nobreakdash-2.
	Events from the chip now arrive at the \mbox{FPGA} with rates of up to two events per \SI{125}{\mega\hertz} \mbox{FPGA} clock cycle and comprise of a \SI{14}{\bit} source neuron address and an \SI{8}{\bit} timestamp \cite{vkarasen20phd}.
	The latter has to be converted to an arrival deadline by adding a modeled axonal delay.
	The lookup table at the source-node now no longer yields a GUID as in \cite{thommes2021bssextoll}, since the destination multicast is not needed with a single chip per \mbox{FPGA}.
	Instead, the lookup now provides a freely remappable destination neuron address.
	
	\subsection{Event Aggregation}

	Figure~\ref{fig:ExperimentSetup} shows the flow of event-streams through the system with ascending timestamps.
	As described in \cite{thommes2021bssextoll}, pulse events are aggregated into larger network packets \cite{rajkumar2008packet} using bucket-buffers.
	The number of events to accumulate is subject to a trade-off between minimizing header-overhead and avoiding congestion when merging packetized event-streams at the destination.
	Also, to avoid timestamp expiration and resulting event-loss, the possible time for aggregation is limited by the modeled axonal delays.\\
	To keep the first prototype implementation simple, the bucket-renaming proposed in \cite{thommes2021bssextoll} and the merging (gray boxes in Figure~\ref{fig:ExperimentSetup}) are not yet realized. 
	Instead, the destination lookup simply yields a bucket-index and the network addresses are statically configured in the buckets.
	In this simplified approach, the required numbers of bucket-units and merge-buffers scale with the number of desired destinations and source-streams per chip.

	\section{Inter-Chip Feed-Forward Neural Network}

	Here we present a technical demonstration of a feed forward neural network spanning two or more HICANN\nobreakdash-X chips.
	A population of neurons on a first chip, driven by external input emits spikes that are then transmitted to a second chip.
	There, they trigger a second population of neurons to answer this firing.
	First measurements using an oscilloscope, attached to analog probing pins yield an overall inter-chip-latency of approximately \SI{8}{\micro\second} (Figure~\ref{fig:CubePhoto}). The membrane-voltage traces in Figure~\ref{fig:ExperimentSetup} have been recorded using the on-chip Membrane Analog to Digital Converter (MADC).
	The synapses have been configured for high technical efficacy in this demo to reliably elicit spikes.

	\section{Summary and Outlook}

	We have presented a first hardware implementation and the according software integration of a low-latency, high-bandwidth communication strategy for multi-chip BSS\nobreakdash-2 systems.
	HICANN\nobreakdash-X chips can be interconnected by transmitting pulse events between \mbox{FPGAs} through a packet-based HPC network.
	This is done using the \mbox{EXTOLL} networking technology. At the conference we present a technical demonstration of these first experiments.

	\begin{acks}
		We thank all present and former members of the Electronic Vision(s) (KIP) and Computer Architecture (ZITI) research groups
		contributing to the BrainScaleS \mbox{EXTOLL} communication hardware and software as well as the \mbox{EXTOLL} company for their technical support with their hardware.
		The research has received funding from the EC Horizon~2020 Framework Programme under Grant Agreements 720270, 785907 and 945539 (HBP),
		the Deutsche Forschungsgemeinschaft (DFG, German Research Foundation) under Germany's Excellence Strategy EXC 2181/1-390900948 (the Heidelberg STRUCTURES Excellence Cluster),
		the German Federal Ministry of Education and Research under grant number 16ES1127 as part of the \textit{Pilotinnovationswettbewerb Energieeffizientes KI-System}, by the Helmholtz Association Initiative and
		Networking Fund (Advanced Computing Architectures, ACA) under project number SO\nobreakdash-092  and
		the Lautenschläger-Forschungspreis 2018 for Karlheinz Meier.
	\end{acks}

	\bibliographystyle{ACM-Reference-Format}
	\bibliography{ExtendedAbstract}

\end{document}